\documentclass[review]{elsarticle}
\usepackage{lineno,hyperref}
\usepackage[english]{babel}
\usepackage[numbers]{natbib}
\usepackage{xcolor}
\usepackage{latexsym,amsmath,amssymb,amsbsy,graphicx,geometry}
\modulolinenumbers[5]

\begin{document}
	
\begin{frontmatter}

\title{Unusual effect of high pressures on phase transformations in Ni$_{62}$Nb$_{38}$ alloy}

\author[kfu,udgu]{B.~N.~Galimzyanov\corref{cor1}}
\cortext[cor1]{Corresponding author}
\ead{bulatgnmail@gmail.com}

\author[kfu]{M.~A.~Doronina}
\ead{maria.doronina.0211@gmail.com}

\author[kfu,udgu]{A.~V.~Mokshin}
\ead{anatolii.mokshin@mail.ru}

\address[kfu]{Kazan Federal University, 420008 Kazan, Russia}
\address[udgu]{Udmurt Federal Research Center of the Ural Branch of the Russian Academy of Sciences, 426067 Izhevsk, Russia}

\begin{abstract}
	Binary Ni$_{62}$Nb$_{38}$ alloy belongs to the unique class of binary off-eutectic systems, which are able to form a bulk glassy state [L. Xia et al., J. Appl. Phys. 99 (2006) 026103]. In the present work, the ($p$, $T$) phase diagram of Ni$_{62}$Nb$_{38}$ alloy was first determined for a wide thermodynamic range with temperatures from $300$\,K to $6000$\,K and with pressures from $1$\,atm to $1.2\times10^7$\,atm. For this thermodynamic range,  elements of the phase diagram such as the liquid-crystal coexistence line and the glass transition line are defined. Our results reveal good agreement between the simulation results and the known experimental values of the liquidus temperature and the glass transition temperature for the isobar $p=1$\,atm. The phase diagram is detailed for pressures greater than $1\times10^{7}$\,atm. For the first time, the phase separation conditions at which the liquid Nb and crystalline Ni phases coexist in the system were determined.
\end{abstract}


\end{frontmatter}

Binary Ni$_{62}$Nb$_{38}$ alloy has a pronounced ability to form bulk metallic glass, which makes this alloy attractive for the manufacture of construction materials~\cite{Jones_DelRio_2021,Xia_Li_2006,Maiorova_Ryltsev_2021,Jeon_Kim_2022,Klumov_Ryltsev_2018}. In the amorphous phase, this alloy has a hardness of approximately $15$\,GPa, a Young's modulus of approximately $230$\,GPa, and a fracture strength of approximately $4$\,GPa. Remarkably, the values of these mechanical characteristics are larger than for other binary bulk metallic glasses, including Ni-based ones~\cite{Lu_Tseng_2021,Xia_Shan_2007,Galimzyanov_Mokshin_2021,Khusnutdinoff_Khairullina_2021}. Ni$_{62}$Nb$_{38}$ bulk metallic glass with a stable amorphous structure was synthesized at the beginning of the 21st century by traditional casting in copper molds~\cite{Xia_Li_2006}. It has been shown that off-eutectic Ni$_{62}$Nb$_{38}$ alloy is the best glass former in comparison with the Ni-Nb system with other ratios of components~\cite{Cheng_Wang_2021,Zeng_Tian_2020,Galimzyanov_Ladyanov_2019}. It is remarkable that this finding is not in agreement with the  empirical rules, according to which bulk metallic glasses can be formed by a multicomponent alloy with a eutectic composition~\cite{Mokshin_Galimzyanov_2020,Alexandrov_Galenko_2020,Baggioli_Zaccone_2021,Zaccone_2021}. 

The structure and mechanical properties of Ni$_{62}$Nb$_{38}$ alloy depend on the thermodynamic ($p$, $T$) conditions in which the alloy is synthesized and/or used. Therefore, to determine properly the functional capabilities of this alloy, it is necessary to know the ($p$, $T$) ranges on the phase diagram at which the system can be in the liquid, crystalline, or glassy phase. Despite Ni$_{62}$Nb$_{38}$ alloy being used for the study of the mechanisms of formation of an amorphous structure~\cite{Xia_Shan_2007,Wen_Zhang_2018,Manna_Pal_2020}, most studies consider this alloy only under standard conditions and in the thermodynamic states with pressures not greater than $1\times10^{3}$\,atm. In addition, the liquidus line $T_l(p)$ and the glass transition line $T_g(p)$ were still unknown for this system. The dependencies $T_l(p)$ and $T_g(p)$ are difficult to determine experimentally because of problems with the realization of cooling and heating protocols at high pressures~\cite{Zhang_Zhou_2022}. Therefore, there are still no experimental data to construct the detailed ($p$, $T$) phase diagram of binary Ni$_{62}$Nb$_{38}$ alloy. On the other hand, a comprehensive study of phase transformations in this system by molecular dynamics simulations became possible after the development of the modified Finnis-Sinclair interatomic interaction potential~\cite{Zhang_Kelton_2016}. This semiempirical potential correctly reproduces the structure of this alloy in the liquid and amorphous states. This is confirmed by good agreement between the simulation results, quantum mechanical calculations, and the findings of X-ray diffraction experiments.

Phase separation in metal alloys into liquid and solid fractions can be observed at high pressures. For example, the phase separation into liquid and crystalline phases or into amorphous and crystalline phases has been observed in Al-based, Cu-based, and Ni-based binary and ternary alloys under high-pressure torsion~\cite{Revesz_2006,Straumal_Korneva_2014}. For the off-eutectic compositions, the phase separation is known and it is a typical effect~\cite{Okamoto_2008}. In the case of Ni$_{62}$Nb$_{38}$ alloy, the phase separation has not been observed previously.

The main aim of the present study is to determine such key elements of the ($p$, $T$) phase diagram of binary Ni$_{62}$Nb$_{38}$ alloy as the liquidus line $T_l=T_l(p)$ and the glass transition line $T_g=T_g(p)$ for a wide range of pressures. Similar studies have not been performed previously for this system. Knowledge of such information makes it possible to determine the crystalline phase region and the region where a bulk metallic glass is formed. We also solve the problem related to determining the ($p$, $T$) conditions at which the phase separation is observed.

In the current study, by means of molecular dynamics simulations we define the ($p$, $T$) phase diagram of binary Ni$_{62}$Nb$_{38}$ alloy covering the temperature range from $300$\,K to $6000$\,K and for pressures up to $1.2\times10^{7}$~atm. The main focus is on the estimation of the liquidus temperature $T_{l}$ and the glass transition temperature $T_{g}$ under various isobaric conditions. We consider the binary alloy with a fixed concentration of atoms: $7229$ Ni atoms and $4435$ Nb atoms are located inside the simulation cubic box with edge length $L\approx63.3$~\AA. The integration of the equations of motion is performed with a time step of $1$~fs. Temperature and pressure are controlled by a thermostat and a barostat according to the Nose-Hoover method~\cite{Thijssen_2007}. The interatomic energies and forces are determined by the semiempirical Finnis-Sinclair potential adapted by Mendelev to reproduce properly the structure and dynamic properties of Ni-Nb alloys in liquid and solid phases~\cite{Zhang_Kelton_2016}. Molecular dynamics simulations were performed with the LAMMPS package~\cite{Plimpton_1995}. Identification of crystalline structures, pair correlation analysis, and visualization of the simulation results were done with the program OVITO~\cite{Stukowski_2009}.

\begin{figure*}[ht]
	\centering
	\includegraphics[width=14.0cm]{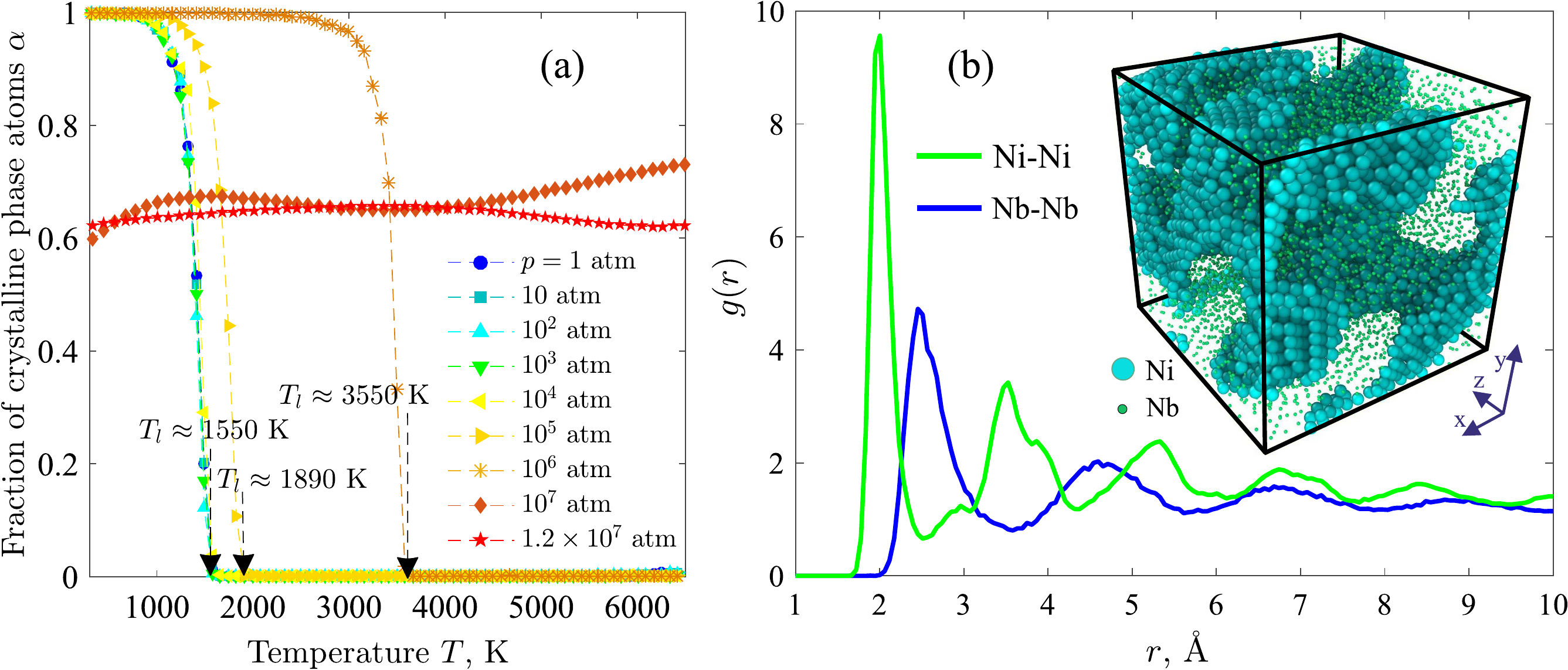}
	\caption{(a) Temperature dependence of the fraction of crystalline phase atoms $\alpha$ for different isobars. Here $\alpha=n/N$, where $n$ is the number of atoms forming the crystalline phase and $N$ is the total number of atoms in the system. (b) Partial pair correlation function $g(r)$ for crystalline Ni and liquid Nb calculated at $6000$\,K and  $1\times10^7$\,atm. The inset shows a snapshot of the system, where Ni forms a branched crystalline structure, while Nb is in the liquid state.}\label{fig_1}
\end{figure*}

The crystalline samples were heated to $6000$\,K at the rate $1\times10^{11}$\,K/s. This fast heating rate is acceptable for molecular dynamics simulations and does not introduce undesirable artifacts such as excessive overlap of atoms, drift of atoms outside the simulation cell, and incorrect control of temperature and pressure~\cite{Zhang_Kelton_2016,Mendelev_Kramer_2010}. The liquidus points on the isobars were determined from the temperature dependencies of the crystalline phase fractions $\alpha$. Figure~\ref{fig_1}(a) shows the temperature dependencies of $\alpha$ for various isobars. The liquidus point corresponds to the temperature $T_l$ at which the system does not contain any crystalline domains and $\alpha$ becomes 0. We found that the liquidus temperature $T_{l}\simeq(1550\pm20)$\,K at  $1$\,atm, which is in excellent agreement with the experimental value $T_{l}^{(Exp)}\simeq1543$\,K~\cite{Lesz_Dercz_2016}. In the pressure range from $1$\,atm to $1\times10^{3}$\,atm, the liquidus temperature increases by only $50$\,K (up to approximately $1600$\,K), which is comparable to the statistical error. A significant increase in the liquidus temperature from approximately $1650$\,K to approximately $3550$\,K is observed at pressures from $1\times10^{4}$\,atm to $1\times10^{6}$\,atm [see Figure~\ref{fig_1}(a)]. For pressures greater than $1\times10^{6}$\,atm, the liquidus temperature goes beyond the considered temperature range. The value of the parameter $\alpha(T)$ fluctuates in the range from $0.6$ to $0.8$ at pressures $1\times10^{7}$\,atm and $1.2\times10^{7}$\,atm because of the coexistence of solid and liquid phases.

The partial pair correlation function $g(r)$ was computed for the system at $6000$\,K and  $1\times10^7$\,atm. As can be seen from Figure~\ref{fig_1}(b), $g(r)$ computed for Ni-Ni and Nb-Nb has features that can be considered as signatures of the so-called phase separation: Ni is in the crystalline phase, while Nb is in the liquid state. The phase separation is directly detected in a snapshot [inset in Figure~\ref{fig_1}(b)], where crystalline Ni domains are embedded in liquid Nb. As we showed earlier~\cite{Galimzyanov_Doronina_2021}, such a redistribution of Ni and Nb atoms is the result of an applied pressure, and not the result of heating.

\begin{figure*}[ht]
	\centering
	\includegraphics[width=14.0cm]{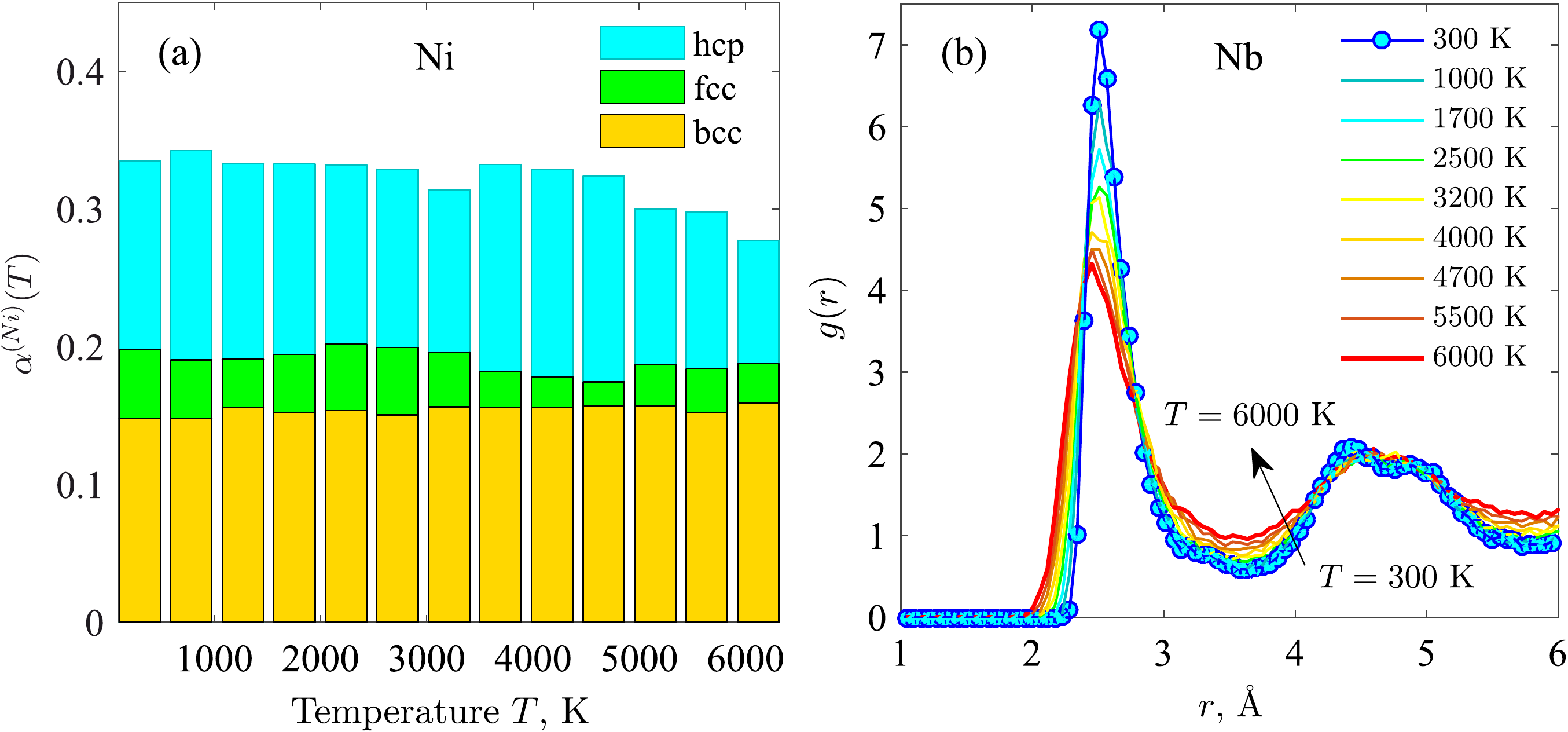}
	\caption{(a) Fraction of Ni atoms in hexagonal close packed (hcp), face-centered cubic (fcc), and body-centered cubic (bcc) crystalline phases as a function of the temperature $T$ at $1\times10^7$\,atm. (b) Pair correlation function for Nb at various temperatures. This figure shows the transition from the amorphous state to the liquid state that occurs when the system is heated from $300$\,K to $6000$\,K at $1\times10^7$\,atm.}\label{fig_2}
\end{figure*}

Figure~\ref{fig_2}(a) shows the fraction of the crystalline Ni atoms forming the hexagonal close packed (hcp), face-centered cubic (fcc), and body-centered cubic (bcc) phases at $1\times10^7$\,atm and in the temperature range from $300$\,K to $6000$\,K. The results indicate the predominance of the hcp phase, while the fraction of the fcc and bcc phase atoms is comparable and does not exceed $40$\%. It is noteworthy that Ni appears in the hcp and fcc modifications as in the case of pure Ni~\cite{Li_Zhang_2013}. Crystalline Ni also appears as an unstable bcc phase. These results agree with the literature data. It was shown earlier in experimental and simulation studies that Ni can exist in the hcp and fcc phases and in the unstable bcc phase at pressures above $100$~GPa~\cite{Boccato_Torchio_2017,Dubrovinsky_Narygina_2007,Tateno_Hirose_2012,Neiva_Oliveira_2016}. Structural rearrangements between the fcc, hcp, and bcc phases are possible at such extremely high pressures, which can occur in solids at terapascal pressures~\cite{Dubrovinsky_Khandarkhaeva_2022,Pickard_Needs_2010}. In Figure~\ref{fig_2}(b), calculated $g(r)$ reveals structural changes in Nb under the considered ($p$, $T$) conditions. We found that Nb is in the amorphous state at  $1\times10^7$\,atm and  $300$\,K. This is confirmed by the presence of the pronounced first maximum and the characteristic splitting of the second maximum of $g(r)$. The amorphous structure is destroyed with increasing temperature, and Nb completely transforms into a liquid state at approximately $4800$\,K. It is remarkable that the similar phase separation in a binary metallic system, at which the liquid/glassy phase of one component and the crystalline phase of other component coexist, has not been observed before at pressures greater than $1\times10^{7}$~atm. In this regard, such separation in Ni$_{62}$Nb$_{38}$ is an unusual effect. In addition, this effect is unusual in that a similar phase separation in binary and ternary metallic systems based on Fe, Al, Cu, and Ni was previously found only for steady-state non-equilibrium conditions when the system is under a moderate pressure (less than $1\times10^{4}$~atm) and torsion load~\cite{Revesz_2006,Straumal_Korneva_2014,Wang_Dai_2001,Liu_Hong_2007,Poryvaev_Polyukhov_2020,Yarullin_Galimzyanov_2020}. For example, the phase separation under compression and heating were observed earlier in  experimental and simulation studies of binary Fe-Cr~\cite{Zhou_Odqvist_2013}, Fe-Cu~\cite{Cheng_Cui_2017}, and Cu-Co~\cite{Bachmaier_Pfaff_2015} alloys as well as multicomponent Fe-Cu-Ge~\cite{Luo_Wang_2017} and LiFePO$_4$~\cite{Yang_Tang_2020}  alloys. Pressure-induced amorphization is most often found in various rocks and minerals formed in the earth entrails~\cite{Pascal_Philippe_1997}. Therefore, the predicted 
impact of pressure on the phase separation in Ni$_{62}$Nb$_{38}$ alloy seems to be correct.

\begin{table}[ht]
	\begin{center}
		\caption{System parameters at different pressures $p$: $T_l$ is the liquidus temperature, $T_g$ is the glass transition temperature.\label{table_1}}
		\begin{tabular}[t]{c|cc}
			\hline\hline
			$p$, atm & $T_l$, K & $T_{g}$, K \\
			\hline
			$1$    & $1550\pm20$ & $1020\pm50$  \\
			$10$   & $1560\pm20$ & $1030\pm50$  \\
			$1\times10^2$ & $1575\pm20$ & $1050\pm60$  \\
			$1\times10^3$ & $1600\pm25$ & $1080\pm60$  \\
			$1\times10^4$ & $1650\pm25$ & $1130\pm60$  \\
			$1\times10^5$ & $1890\pm30$ & $1220\pm70$  \\
			$1\times10^6$ & $3550\pm50$ & $1780\pm80$  \\
			\hline\hline
		\end{tabular}	
	\end{center}
\end{table}

\begin{figure*}[ht]
	\centering
	\includegraphics[width=16.0cm]{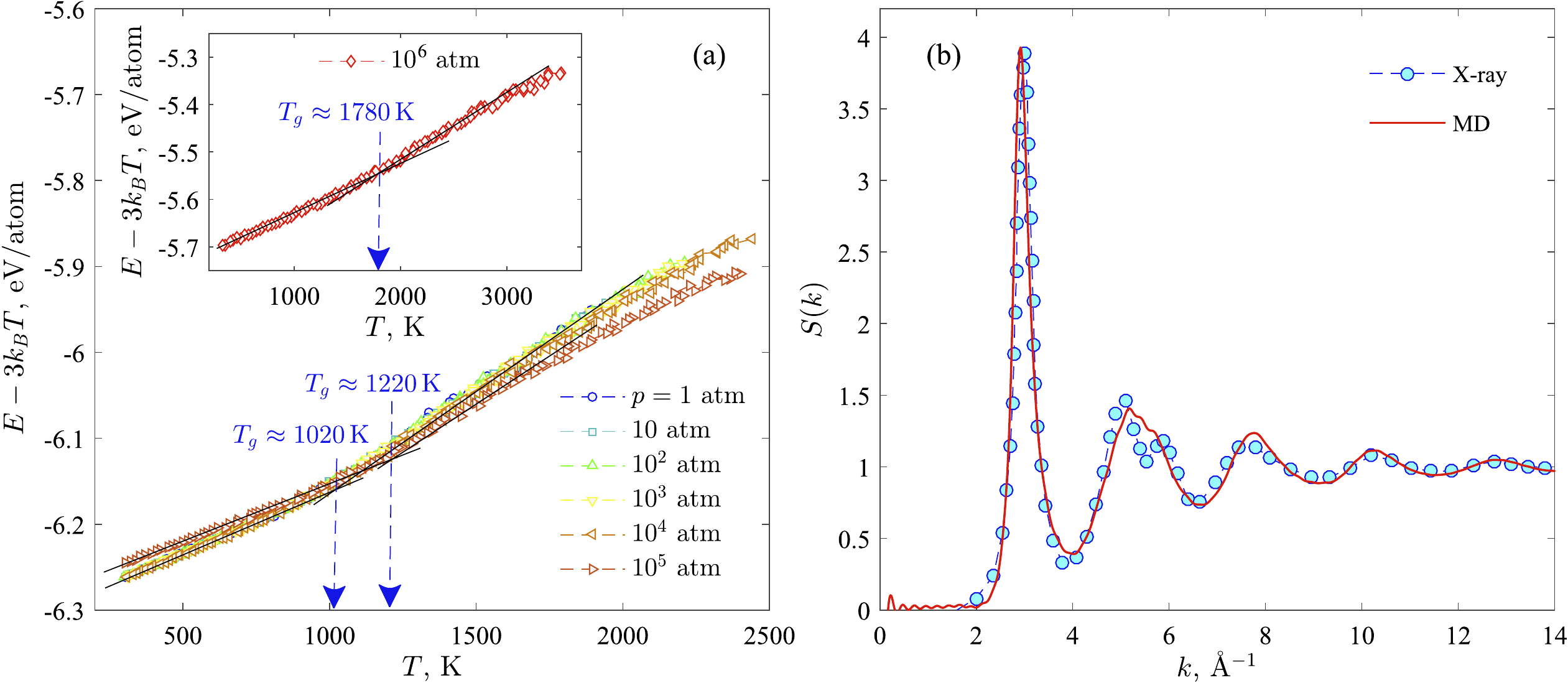}
	\caption{(a) Difference $E-3k_{B}T$ between potential and kinetic energies of the system as a function of the temperature $T$ at different isobars. (b) Structure factor $S(k)$ calculated through molecular dynamics simulations (MD) at $300$\,K and  $1$\,atm compared with X-ray diffraction data~\cite{Zhang_Kelton_2016,Mauro_Johnson_2013}.}\label{fig_3}
\end{figure*}

The glass transition conditions were determined during rapid cooling of equilibrium liquid melt. A liquid system with temperature $1.5T_{l}$ was cooled at the rate $1\times10^{12}$\,K/s to $300$\,K on various isobars. The glass transition temperature $T_{g}$ is determined from the change in the difference between the potential energy $E$ and the kinetic energy $3k_{B}T$ at decreasing temperature $T$. Figure~\ref{fig_3}(a) shows the temperature dependence of the difference $E-3k_{B}T$ computed at different pressures. It can be seen from Figure~\ref{fig_3}(a) that these dependencies contain two regimes: (i) the high-temperature regime, corresponding to the liquid state, in which the energy decreases rapidly; and (ii) the low-temperature regime, corresponding to the frozen state, where the energy decreases slowly. The boundary between these regimes on the temperature scale directly corresponds to the glass transition temperature $T_{g}$ [see Figure~\ref{fig_3}(a)]. The size of this boundary determines the error interval in the found values of $T_{g}$ (see Table~\ref{table_1}).

In Figure~\ref{fig_3}(b), the computed structure factor $S(k)$ of amorphous Ni$_{62}$Nb$_{38}$ alloy is compared with the X-ray diffraction data~\cite{Zhang_Kelton_2016,Mauro_Johnson_2013}. The simulation data and the experimental data were obtained under identical thermodynamic conditions. As can be seen from Figure~\ref{fig_3}(b), there is excellent agreement between the simulation data and the X-ray diffraction data except in the region of wave numbers corresponding to the second peak of the static structure factor $S(k)$. Further,  the glass transition temperature $T_{g}\simeq1020$\,K at  $1$\,atm is close to the known experimental value $T_g^{(Exp)}=891 $\,K~\cite{Mauro_Johnson_2013}. As is known, the faster the cooling rate, the greater the glass transition temperature $T_g$~\cite{Ninarello_2017,Hutchinson_2009}. In the present work, the cooling rate is $1\times10^{12}$\,K/s, while in Ref.~\cite{Mauro_Johnson_2013} the reported results  were obtained for the case of cooling at a rate of approximately $10$\,K/s. Therefore, it is quite obvious that the obtained glass transition temperature is approximately $130$\,K larger than the experimental glass transition temperature. 
\begin{figure*}[ht]
	\centering
	\includegraphics[width=11.0cm]{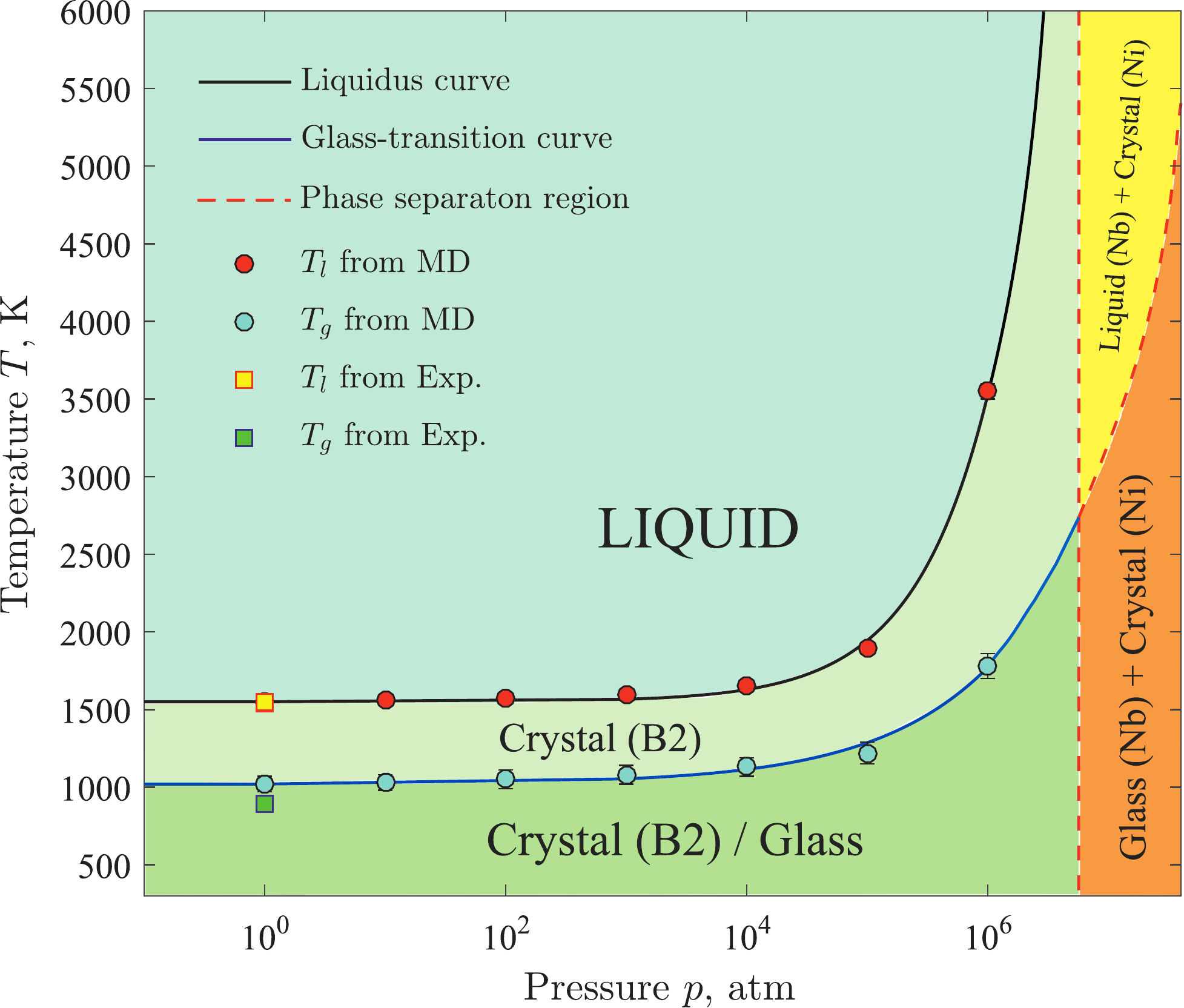}
	\caption{($p$, $T$) phase diagram of Ni$_{62}$Nb$_{38}$ alloy. The pressure is presented on a logarithmic scale. The liquidus line (black) and the glass transition line (blue) were obtained with use of Eq.~(\ref{eq_melt_line}) and Eq.~(\ref{eq_glass_line}), respectively. The circles indicate the calculated temperatures $T_{l}$ and $T_{g}$ taken from Table~\ref{table_1}. The squares denote the experimentally measured temperatures $T_{l}$ and $T_{g}$ at  $1$\,atm~\cite{Lesz_Dercz_2016,Mauro_Johnson_2013}. The dotted red lines indicate the phase separation regions at pressures $p\geq1\times10^{7}$\,atm. Exp., experiments; MD, molecular dynamics simulations.}\label{fig_4}
\end{figure*}

Figure~\ref{fig_4} shows the ($p$, $T$) phase diagram obtained for Ni$_{62}$Nb$_{38}$ alloy. In this diagram, the liquid-solid equilibrium lines and the boundaries of the phase separation regions are depicted. We found that the pressure dependence of the calculated liquidus temperature $T_{l}$ is accurately reproduced by the well-known Simon-Glatzel (SG) empirical equation~\cite{Schlosser_Vinet_1989,Errandonea_2010}:
\begin{equation}\label{eq_melt_line}
T_{l}(p)=T_{l0}\left(1+\beta p\right)^{\xi}.
\end{equation} 
This equation has a simple form and contains a minimum number of adjustable parameters compared with other empirical equations and its modifications. In Eq.~(\ref{eq_melt_line}), $T_{l0}\simeq1550$\,K is the liquidus temperature at $p=1$\,atm. The adjustable quantities $\beta$ and $\xi$ are related to the Gr\"{u}neisen parameter~\cite{Schlosser_Vinet_1989}. The best agreement between the simulation data and Eq.~(\ref{eq_melt_line}) was obtained with $\beta\simeq(7.5\pm2.0)\times10^{-6}$\,atm$^{-1}$ and $\xi\simeq 0.38\pm0.4$. These values are comparable with known literature data obtained for metals. For example, the melting line of pure nickel is reproduced by Eq.~(\ref{eq_melt_line}) with $\beta\simeq1\times10^{-6}$\,atm$^{-1}$ and $\xi\simeq0.45$ (see Table I in Ref.~\cite{Babb_1963}), which are close to the found values for Ni$_{62}$Nb$_{38}$ alloy.

As can be seen from Figure~\ref{fig_4}, the correspondence between the temperature $T$ and the pressure $p$ for the glass transition is similar to the liquidus line. Therefore, it is also convenient to interpolate the  data obtained for the glass transition by the SG-type equation~\cite{Schlosser_Vinet_1989,Kaminski_Pawlus_2012} 
\begin{equation}\label{eq_glass_line}
T_{g}(p)=T_{g0}\left(1+\frac{p}{\Pi}\right)^{1/b}.
\end{equation} 
Note that Eq.~(\ref{eq_glass_line}) is identical to the phenomenological Andersson-Andersson equation~\cite{Andersson_Andersson_1998}. Here, $T_{g0}\simeq1020$\,K is the glass transition temperature at  $p=1$\,atm.  $\Pi$ and $b$ are adjustable and depend on the system type. $\Pi$ has the dimension of pressure, while the dimensionless parameter $b$ is defined through the volume expansion coefficient and the specific heat capacity of the system~\cite{Kaminski_Pawlus_2012}. For the  binary alloy considered, we found that $\Pi\simeq(33\pm10)\times10^{3}$\,atm and $b\simeq 6.25\pm1.0$. The $T_g(p)$ dependence obtained for Ni$_{62}$Nb$_{38}$ alloy by Eq.~(\ref{eq_glass_line}) is correct and similar in shape to the experimentally measured glass transition lines of real glass-forming systems~\cite{Drozd_Rzoska_2007,Buchholz_Paul_2002}. Note that the obtained dependencies $T_l(p)$ and $T_g (p)$ are in agreement with previous results~\cite{Schlosser_Vinet_1989,Drozd_Rzoska_2007}. Namely, the found values of the exponents in SG-type equations (\ref{eq_melt_line}) and (\ref{eq_glass_line}) belong to the interval ($0$; $1$), which is valid for various types of liquid (including metallic ones) with pronounced glass-forming ability~\cite{Babb_1963}.

In summary, the ($p$, $T$) phase diagram of Ni$_{62}$Nb$_{38}$ alloy  was first determined for a wide thermodynamic range. The liquidus temperature $T_l$ and the glass transition temperature $T_g$ as a function of the pressure $p$ were determined. We have shown that obtained dependencies $T_l(p)$ and $T_g(p)$ are reproduced by the well-known SG-type empirical equations. The phase separation is observed at pressures above $1\times10^{7}$\,atm: Nb is in the liquid state, while Ni forms a percolating crystal structure. This result is of great fundamental importance since it shows that the pressure is one of the main thermodynamic parameters that allows one to control the phase transformations in Ni$_{62}$Nb$_{38}$ alloy. Therefore, the results of the present work could be a starting point for study of the thermodynamics of Ni-Nb systems with various concentrations of Ni atoms. 

Funding: This work was supported by the Russian Science Foundation (project no. 19-12-00022-P).

Supplementary material: Computational details.

\bibliographystyle{unsrt}

\end{document}